# Hierarchy of events in protein folding: beyond the Gō model

Ludovico Sutto[1], Guido Tiana[1] and Ricardo A. Broglia[1,2]

[1]Department of Physics, University of Milano and INFN, via Celoria 16, 20133, Italy
[2]The Niels Bohr Institute, Bledgamsvej 17, 2100 Copenhagen, Denmark

(21[th] December 2005)

Corresponding Author:

Guido Tiana
Department of Physics
via Celoria 16
20133 Milano
Italy
tel: +39-02-50317221
fax: +39-02-50317487
email: tiana@mi.infn.it

Total number of manuscript pages: 38

Number of tables: 4

Number of figures: 14

*Abstract*

Simplified Gō models, where only native contacts interact favorably, have proven useful to characterize some aspects of the folding of small proteins. The success of these models is limited by the fact that all residues interact in the same way, so that the folding features of a protein are determined only by the geometry of its native conformation. We present an extended version of a $C_\alpha$-based Gō model where different residues interact with different energies. The model is used to calculate the thermodynamics of three small proteins (Protein G, Src-SH3 and CI2) and the effect of mutations ($\Delta\Delta G_{U\text{-}N}$, $\Delta\Delta G_{\ddagger\text{-}N}$, $\Delta\Delta G_{\ddagger\text{-}U}$ and $\phi$-values) on the wild-type sequence. The model allows to investigate some of the most controversial areas in protein folding such as its earliest stages, a subject which has lately received particular attention. The picture which emerges for the three proteins under study is that of a hierarchical process, where local elementary structures (LES) (not necessarily coincident with elements of secondary structure) are formed at the early stages of the folding and drive the protein, through the transition state and the post-critical folding nucleus (FN), resulting from the docking of the LES, to the native conformation.

INTRODUCTION

The simulation of the folding of proteins by means of realistic models, like Gromacs (Berendsen et al. 1995) or Amber (Pearlman et al. 1995), is still computationally out of reach. Even more hopeless looks the possibility of obtaining, from such simulations, quantities reflecting the thermodynamical and kinetical properties of the folding process, in keeping with the fact that to acquire this knowledge with meaningful statistics requires the calculation of thousands of folding and unfolding events.

In lack of soluble realistic models, use is commonly made of simplified descriptions of the protein. Among them, a widely used approach is provided by the Gō model (Gō 1983), which makes use of a potential function based on the knowledge of the native structure of the protein. This potential is, as a rule, the sum of two-body terms contributing with -1 if a native contact is formed and zero otherwise. Within this framework, different degrees of approximation can be made concerning the description of the amino acids ranging from all atoms- to $C_\alpha$-representations.

These models have the virtue of making, by definition, the native state to be the global energy minimum of the system aside from making computationally feasible the description of the folding process. On the other hand, they neglect the chemical properties of the different types of amino acids, treating all of them on equal footing. While these models describe reasonably well the entropy of the chain, they simplify drastically concerning the energy of the system. This fact poses a number of limitations in the usefulness of Gō models. Most noticeably, within this model, the properties of a protein are solely determined by its geometry. This is at variance with the fact that proteins displaying the same (native) structure but different sequences can display different stability (Guerois et al. 2002), folding rates (Grantcharova et al. 1998, Martinez et al. 1998) and overall folding features (Khan et al. 2003). The hierarchy of events which one gets from Gō model calculations would be dependent only on the geometric separation of residues along the chain (i.e., closer pairs are formed first), while we expect it to depend also on the interaction energy between residues.

To overcome some of these problems, we propose a modified Gō model, where native contacts interact with a pair-dependent potential, while non-native contacts display only nonspecific core repulsion, as in standard Gō models. The parameters which control the pair potential are calculated for each protein from the measurement of the destabilizing effect mutations have on the native conformation (experimental $\Delta\Delta G_{U-N}$ values).

The Gō model based on the experimental $\Delta\Delta G_{U-N}$ values is then applied to three small proteins (Protein G (56 residues), Src-SH3 (60 residues) and CI2 (64 residues)) for which a consistent amount of mutational data is available (see Fig. 1). The effectiveness of the model is first tested comparing the properties of the transition state (namely, $\Delta\Delta G_{U\text{-}\ddagger}$ and $\Delta\Delta G_{N\text{-}\ddagger}$), calculated within the model following the same procedure used to extract those quantities from *in vitro* experimental data.

We then study the hierarchy of structure formation during the folding process, starting from random conformations. We find, for the three proteins under consideration, a well defined sequence of events, where first stable local elementary structures (LES) (Broglia et al. 2004) are formed, involving a small set (2-4) of *fragments* of the protein stabilized by some of the most strongly interacting amino acids. When these structures build their mutual native contacts, that is the post critical FN (Abkevich et al. 1994), the transition state is overcome and the protein folds on very short call to its native conformation. The post-critical FN is defined in Abkevich et al. 1994 as "the minimal sized fragment of the new phase [i.e. the folded phase] that inevitably grows further to the new phase". Consequently, it lies beyond the transition state, its formation being a sufficient and necessary condition for folding. The LES can be defined as the fragments of consecutive residues which build out the post-critical FN (Broglia et al. 2004). Note that the definition we have used of the (post-critical) FN, as the minimum set of native contacts which brings the system over the highest barrier of the free energy associated with the folding process, is not inconsistent with the definition of Abkevich et al. 1994. This dynamical picture centered around the LES and the FN is somewhat complementary to the chemical view of protein folding, which emphasizes the transition state. The free-energy-defined transition state, arising from the interpretation of protein folding as a chemical reaction, can be structurally evanescent and suffers the lack of a set of grounded reaction coordinates. The FN, on the other hand, is not located precisely on the free-energy landscape (one only knows that it is in the native basin) but it is structurally well defined and reflects directly the hierarchy of folding events.

The idea that a hierarchical set of events leads proteins to their native state, avoiding a time-consuming search through conformational space, is well established. Ptitsyn and Rashin observed a hierarchical pathway in the folding of Mb (Ptitsyn et al. 1975). Lesk and Rose identified the units building the folding hierarchy of Mb and RNase on the base of geometric arguments (Lesk et al. 1981), deriving the complete tree of events which lead these proteins to the native state. All these works describe a framework where small units composed of few consecutive amino acids build larger units which, in turn, build even larger ones, which eventually involve the whole protein. This mechanism has been suggested to take place also for proteins which apparently follow a non-hierarchic route, such as those described by the nucleation-condensation model (Baldwin et al. 1999a). The kinetic advantage of this mechanism (as compared to non hierarchical scenarios) is that, at each level of the hierarchy, only a limited search in conformation space is needed for the smaller units to coalesce into the larger units belonging to the following level (Panchenko et al. 1995) of organization.

An analytical model developed by Hansen and coworkers, although lacking of molecular detail, has shown that a hierarchical folding mechanism is not incompatible with a cooperative folding transition, like that displayed by two-state folders (Hansen et al. 1998).

Making use of a simplified model where a protein is described as a chain of beads on a cubic lattice, interacting through a contact potential, it was shown (Tiana et al. 2001, Broglia et al. 2001a) that designed sequences fold by building, at the very beginning of the dynamical process, LES, composed of few consecutive amino acids and stabilized by the most attractive contact matrix elements. The folding time is essentially determined by the time needed by the LES to dock, thus forming the FN. After this event has taken place the remaining residues fold very rapidly, due to the strongly reduced size of the conformational space remained to them. Moreover, exploiting the hierarchical character of the folding mechanism, it has been possible to successfully predict the native conformation of lattice-model designed proteins from the knowledge of the sequence and of the potential function alone (Broglia et al. 2001b). In other words, to solve the protein folding problem of designed proteins in the lattice.

In concluding this section one should remember that the proteins under study (Protein G, Src-SH3 and CI2) are domains of real proteins. In particular, protein G corresponds to the binding domain of the streptococcal bacterium to which the mammalian immunoglobulin IgG attaches to signal the immune system of the host the presence of an intruder. To which extent the rest of the protein may affect the folding of the corresponding domains is not known in detail. Because the picture of the folding process based on the concept of LES emerged from the study of designed (lattice) model proteins (Tiana et al. 2001, Broglia et al. 2001a, Broglia et al. 2001b) as well as of real proteins (like e.g. the HIV-1-PR (Broglia et al. 2005, Levy et al. 2004)), it may not be exactly applicable to the proteins under consideration. Anyway, it will be shown that the hierarchical picture of the folding process seems to be quite appropriate also to describe the folding of the G, SH3 and CI2 domains.

RESULTS

**Derivation of the potential**

To be able to carry out extensive simulations the atomic structure of the amino acids is neglected and each monomer is substituted by a hard sphere centred at position $C_\alpha$ (cf. *Methods* for details about the geometric aspects of the model).

The choice of the potential function is a key issue of the present work. Current force fields based on the chemical properties of amino acids meet, as a rule, difficulties in predicting detail properties of natural proteins (Lazaridis et al. 1999). For this reason, use is made of a potential based on the native conformation of the protein and on the experimental free energy changes upon mutation which gives an energy $B_{ij}$ to every pair of residues *i,j* building a native contact. That is,

$$U(\{r_i\}) = \sum_{i+2<j} B_{ij} \cdot \theta(R-|r_i-r_j|) \cdot \theta(R-|r_i^N-r_j^N|) +$$
$$+ \epsilon \cdot \theta(0.99|r_i^N-r_j^N|-|r_i-r_j|) \cdot \theta(R-|r_i^N-r_j^N|) + \quad (1)$$
$$+ \epsilon \cdot \theta(R-|r_i-r_j|) \cdot \theta(|r_i^N-r_j^N|-R),$$

where $r_i$ is the position of the *i*th residue, $B_{ij}$ is the matrix element giving the interaction between the *i*th and *j*th residue, $\theta(x)$ is the Heaviside step function which assumes the value *1* if *x>0* and *0* otherwise, $\{r_i^N\}$ are the coordinates of the crystallographic native conformation, *R* is the interaction range, while $\varepsilon$ is a hard-core repulsion energy set to *100 $k_BT$* (see *Methods*).

The numerical values of the matrix $B_{ij}$ are calculated from the experimental values of the change in free energy $\Delta\Delta G_{U-N}$ of the native state with respect to the unfolded state upon mutations. Assuming that 1) the entropy of the native state does not change after the mutation, 2) the effect of the mutation is to make the interaction energy of the mutated amino acid negligible, and 3) the interaction energy $B_{ij}$ cannot grow above *max(-$\Delta\Delta G_{U-N}$(i), -$\Delta\Delta G_{U-N}$(j))*, one can write

$$\sum_{j=1}^{L} B_{ij} \cdot \theta(R-|r_i^N-r_j^N|) = \Delta\Delta G_{U-N}(i), \quad (2)$$

If all $\Delta\Delta G_{U-N}(i)$ were known, Eq. (2) provides a set of *L* linear equations (*L* being the number of the amino acids which form the protein and thus determine its length) in the variables $B_{ij}$, whose number is $\gamma L/2$ ($\gamma$ being the average number of contacts that each amino acid builds in the native conformation). Since usually $\gamma > 2$, the number of variables exceeds that of equations and one can determine the quantities $B_{ij}$ except for $\gamma L/2$-*L* free parameters. In general, this allows the $B_{ij}$ to range over all real numbers. On the other hand, the assumption 3) constrains the values of $B_{ij}$ from above, making the range of uncertainty much smaller. The rationale behind this assumption is the "principle of minimal frustration" according to which protein sequences have evolved over millions of years decreasing, to the maximum extent, the energetic contradictions among amino acids (Bryngelson et al. 1987). In other words, a residue building, for example, two contacts could, without the constraint, increase the energy of one of them by an arbitrary value, while decreasing the other by the same amount. This would increase the frustration of the system, due to the presence in the native state of a large repulsive interaction. The effect of the constraint 3) stated above is to prevent an amino acid with a given value of $\Delta\Delta G_{U-N}(i)$ from giving rise to energies $B_{ij}$ which are very large in absolute value and which through strong cancellation, lead still to the right native conformation energy but at a strongly reduced stability. Note that numerical values of $\Delta\Delta G_{U-N}(i)$ are often not available for all residues, giving rise to an even larger number of free parameters. To circumvent this obstacle, we have stochastically solved the set of equations producing a statistically representative set of solutions, the parameters $B_{ij}$ used corresponds to the average of these solutions.

While we shall soleley use this value of $B_{ij}$ in all the calculation presented in this paper, it is of interest to compare these results with those obtained making use of GROMACS. This is done in Fig. 2, where we display the energy ($=\Sigma_j B_{ij}$) associated with each residue $i$ of the three proteins under study in their native conformation. While marked deviations between the two sets of energies are observed, the overall trend of the empirical values is in overall agreement with the values calculated with GROMACS. We note that the energies associated with the empirical $B_{ij}$ values deduced from $\Delta\Delta G_{U-N}$ measurements display a smoother behaviour than that associated with the realistic potential (GROMACS). This difference indicates that the prescription based on Eq. (2) and conditions 1)-3) above is likely to distribute much too uniformly the contact energies associated with the mutated amino acids, among those residues which have not been subject to mutations. Consequently, one would expect that the features observed with the present generalized Gō model are less marked than those one would obtain carrying out a (not possible to date) full dynamical simulation with GROMACS.

**Model calculations of experimentally accessible quantities**

In order to validate the model, we have investigated a number of quantities associated with the native (N), transition (‡) and unfolded (U) state of proteins and compared them with the experimental data.

In the case of Protein G (cf. Fig. 1(a)) the experimental $\Delta\Delta G_{U-N}$, $\Delta\Delta G_{‡-N}$ and $\Delta\Delta G_{‡-U}$ are available for 26 out of 56 residues (McCallister et al. 2000, Park et al. 1999). The associated $B_{ij}$ matrix has an average value $<E>=-0.56$ and a standard deviation $\sigma=0.44$ expressed in *kcal/mol*, ($k_B=1$). Making use of these energies, Monte Carlo simulations have been performed in order to elucidate the thermodynamics of the protein. In particular, we have studied the parameters $q_E$, which is the fraction of native energy $E/E_N$ of a given conformation, and the distance root mean square deviation $dRMSD=\sqrt{\frac{1}{N(N-1)}\sum_{i,j}(d_{ij}-d_{ij}^N)^2}$, where $d_{ij}^N$ is the relative distance between amino acids $i$ and $j$ in the native conformation, while $N$ is the total number of amino acids. The equilibrium probability $p(q_E,dRMSD)$ is displayed in Fig. 3(a), as a function of $q_E$ and $dRMSD$. The probability shows two peaks, one associated with the native state (high $q_E$ and low $dRMSD$) and the other associated with the unfolded state (low $q_E$ and high $dRMSD$).

The folding temperature for protein G, at which the population of the native state is equal to that of the unfolded state, is $T_f$=0.37 (in *kcal/mol*, with $k_B$=1). Note that this is, in absolute terms, an unphysically low temperature (*45 K*), which reflects the approximations introduced in the model (i.e., consider only the $C_\alpha$, neglect of the solvent, etc.). In fact, in a model controlled by less degrees of freedom than the actual number of degrees of freedom of the system under consideration, one expects that entropy changes are more abrupt than those taking place in the real system. Since temperature is defined as the inverse derivative of entropy with respect to energy, the increased steepness of entropy results in lower temperatures. For this reason we will express temperatures relatively to $T_f$.

Within this context, the conversion of Monte Carlo steps (MCs) to time used below may resent the reduced number of degrees of freedom explicitly treated by the model (for more details see *Methods*).

Defining operatively a mutation as a switch off of all the native contacts that the mutated residue displays in the native conformation, it is possible to calculate the effect of mutations on the folding and unfolding of the protein (see *Methods*). A comparison between model-calculated and experimental values of the variation of free energy $\Delta\Delta G_{\ddagger-N}$, $\Delta\Delta G_{U-N}$ and φ-values is displayed in Fig. 4. The correlation coefficients $r$ and the root mean square deviation $\sigma$ between theoretical and experimental values are listed in Table 1 (first row) and indicate an acceptable degree of performance of the model. We have repeated the above calculations employing a model with the same geometry but with a pure Gō potential (i.e., $B_{ij}$=-1 for all native pairs). The results are also shown in Table 1 (second row) and indicate a decrease of the correlation coefficient and an increase of the root mean square deviation as compared to the corresponding values of the pair specific potential used in the extended Gō model, with the exception of $\Delta\Delta G_{\ddagger-U}$.

The thermodynamics calculations have been repeated within the framework of the present model also for Src-SH3 and CI2 (cf. Figs. 1(b) and 1(c) ), calculating the interaction matrices $B_{ij}$ from the experimental $\Delta\Delta G_{U-N}$ (Riddle et al. 1999, Itzhaki et al. 1995) which are known for 37 out of 60 residues and 33 out of 64 residues respectively. The two matrices are characterised respectively by an average energy *<E>=-0.29* and *<E>=-0.53* and standard deviation *σ=0.37* and *σ=0.47* expressed in *kcal/mol* (with $k_B$=1). Making use of the matrix elements $B_{ij}$ the energies $\Sigma_j B_{ij}$ associated with each amino acid $i$ in the native conformation were calculated. The resulting values are displayed in Figs. 2(b) and 2(c) in comparison with the prediction of the software GROMACS. The agreement in the case of the Src-SH3 protein is excellent, while in the case of protein CI2 the same comments made in connection with protein G (see Fig. 2(a)) applies (overall reproduction of trend but conspicuous local deviations).

The equilibrium probabilities $p(q_E,dRMSD)$ for these two proteins are displayed in Figs. 3(b) and 3(c) respectively and show a two-state behaviour for Src-SH3. On the other hand the peaks for CI2 are quite broad, much more than in the case of Src-SH3, suggesting a less cooperative folding process, results which are compatible with the experimental data. The folding temperatures associated with Src-SH3 and CI2 proteins are $T_f=0.21$ and $T_f=0.32$ *kcal/mol*, respectively. The results concerning mutations are displayed in Figs. 5 and 6, while the correlations with the experimental data are listed in Table 1 (third and fourth rows). The data indicate that, while the ability of the model in predicting the $\Delta\Delta G_{\ddagger\text{-}N}$ and $\Delta\Delta G_{U\text{-}N}$ is still acceptable, the $\Delta\Delta G_{\ddagger\text{-}U}$ and $\phi$-values are completely missed. The meaning of these results will be faced in the *Discussion* section.

**Folding events**

The interest of the model is that it allows for the study of the early events of folding, a subject of particular interest in view also of recent experimental developments (Religa et al. 2005). We have performed 200 dynamical simulations for each of the three proteins, starting from random conformations recording the mean value $[q_E](t)$ of $q_E$ and the probability of formation of each native contact $p_{i\text{-}j}(t)$ as a function of time.

*Protein G*

The results for Protein G at $T/T_f=0.54$ are displayed in Figs. 7(a) and 8. The curve $[q_E](t)$ is well fitted by the sum of two exponential, of characteristic times of the order of *10 ns* and *1 µs*, respectively. Note that continuous-flow experiments (McCallister et al. 2000) performed at acidic conditions can be fitted by two exponentials of characteristic times *300 µs* and *2 ms*, respectively, while at neutral pH only one exponential is observed at sub-millisecond time scale. Consequently the results of our model, whose parameters are obtained from $\Delta\Delta G_{U\text{-}N}$ measured at neutral pH, are not incompatible with the folding kinetics, since the nanosecond events fall in any case under the instrumental dead-time.

According to the $p_{i\text{-}j}(t)$ (cf. Figs. 7(a), 8, 9 and 10), the first structures which are formed within the first nanosecond (=$10^4 MCs$) are the second hairpin (residues 41-56) and the most local contacts within the α-helix. The former starts from the turn (contacts 46-49 and 47-50) and closes up until contact 44-53, but also up to contact 39-56 (see Fig.9(II)). All of these early contacts are very stable (cf. Fig. 7(a)-A) and their dynamics is single-exponential (cf. Fig. 8, dark grey curve $p_{44\text{-}53}(t)$ ). The folding process then proceeds with the formation of the first hairpin (residues 1-20) and the further stabilization of the second hairpin which ends after *10 ns* (contact 41-54). Whereas the formation of the second hairpin resemble the closure of a zip (see the colour gradient associated with strands β3-β4 in Fig. 7(a)-B as well as Fig. 9) this behaviour is not seen for the first hairpin (strands β1-β2 in Fig. 7(a)-B). The formation of the full α-helix and the formation of the contacts between the two hairpins (strands β1-β4 in Fig. 7(a)) take place on a much longer time scale than the previous events (i.e., microseconds) following a non-exponential dynamics (cf. Fig. 8, where the dark grey curve $p_{5-52}(t)$ represents a contact between strands β1-β4). As expected, the most stable contacts are the ones within the hairpins and within the α-helix, while the less stable are those across the hairpins (cf. Fig 7(a)-A).

These results can also be interpreted in terms of three LES $S_1$(4-9), $S_2$(41-46) and $S_3$(49-54) which essentially control the folding process. Because $S_2$ and $S_3$ dock very fast (~$10^3$-$10^4$ MCs, 0.1-1ns) giving essentially rise to a closed LES (second hairpin $S_{II}$(41-54)), we can equally well describe the folding of the G protein in terms of $S_I$($\equiv S_1$ ; open LES) and $S_{II}$. In fact, the docking of $S_I$ and $S_{II}$ lead to the post critical folding nucleus (FN). This event takes place in about *15M MCs* (≈*1-1.5μs*) (dRMSD~5.3Å) for the trajectory shown in Fig. 9, after which the protein folds shortly after (FPT≈ *20M MCs, 2μs*) (see also Fig. 10).

It is known that LES structures are, as a rule, stabilized by strongly interacting highly conserved amino acids. Mutation of hot amino acids (typically 8-10% of all the amino acids) have the largest impact on the ability the protein has to fold. Mutation of warm amino acids (typically 17-20% of all amino acids) although not leading as a rule to denaturation, can slow to a certain extent the folding process and eventually slightly alter the native conformation (Broglia et al. 2004).

From Fig. 4 it is seen that the five (≈0.08×56) amino acids occupying sites 21, 41, 45, 52 and 54 have the largest impact on the ability the protein has to fold (see Fig.4), and thus can be viewed as hot amino acids (see e.g. Broglia et al. 2004). Within this picture the ten (≈0.17×56) amino acids 6, 7, 20, 22, 31, 34, 35, 39, 46 and 51 can be viewed as warm amino acids. Note that $S_{II}$ is built out by all (but one, namely amino acid 21) hot amino acids, while $S_I$ contains two warm amino acids.

*Protein Src-SH3*

Analogous simulations have been performed for the Src-SH3 protein at *T/T$_f$=0.86* giving a two exponential fit of the curve $[q_E](t)$ with characteristic time *13 ns* ($10^5$MCs) and *2 μs* ($2x10^7$MCs) (data not shown). From the contact map of stability and formation time displayed in Fig. 7(b) as well as from Figs. 11 and 12 it emerges that the first contacts which get formed are the most local ones within the distal hairpin (contacts 41-44 and 42-45 forms in *0.1 ns, $10^3$MCs*) and within the 3$_{10}$-helix (contact 52-55). In the same short time scale contact 21-24 of the diverging turn is fully stabilized. The folding then proceeds with the full formation of the distal hairpin which reaches its maximum stability when contact 36-51 forms (*5-10 ns, $10^5$MCs*). After this, in a time scale of the order of *30-50 ns*, the RT-loop and the contacts between strands β1 and β2 get formed. These contacts, together with the distal hairpin, give rise to the most stable structures of the protein (cf Fig.7(b)-A). The next events take place on a much longer time scale (*≈0.25 μs, 2.5M MCs*) and involves the docking of strand β2 to β3, strand β4 to the RT-loop and finally strand β1 to β5. These last events occur almost simultaneously and lead the protein to its compact native conformation.

The folding of the Src-SH3, associated with a FPT of $\tau_f$≈2 *μs (20M MCs)* can also be interpreted in terms of the LES $S_1$(3-10), $S_2$(18-26), $S_3$(36-44) and $S_4$(47-51) (local elementary structures which, used as peptides (p-$S_i$ (i=1,2,3,4)) in the typical ratio 3:1 (peptide-protein), strongly inhibits the folding of the protein (Broglia et al. 2004, Broglia et al. 2005)). These LES contain all of the hot (10,18,20,24,26,50) and warm (5,7,23,38,41,44$^*$,48$^*$,49,51) amino acids, as emerged from the experimental $\Delta\Delta G_{U-N}$ values (see Fig. 5) (the site marked with an asterisk being amino acids intermediate between hot and warm, i.e. warm-hot amino acids). The formation of the native bonds between $S_3$ and $S_4$ gives rise, very early in the folding process (*5-10ns, 50k-100k MCs* see Figs. 11(II) and 12) to the distal hairpin (see Table 3), structure which can be viewed as a closed LES. Somewhat later, but still at the very beginning of the folding process (*30-50ns, 0.3-0.5M MCs* see Figs. 11(III) and 12), $S_1$ and $S_2$ form their local contacts leading to the formation of the RT-loop and the diverging turn, structure which can be viewed as the second closed LES. Most of the folding time is spent by the two closed LES, in exploring conformational space in search of the correct relative distance and orientation leading to their docking. This event which takes place after approximately *1.6-1.8μs* (*16-18M MCs*) (see Figs. 11(IV) and 12) gives rise to the (post critical) FN. Shortly after the remaining amino acids find their native position and the protein reaches for the first time the native conformation (*2μs*).

*Protein CI2*

Finally, analysing the contact map displayed in Fig. 7(c) and Figs. 13 and 14 and the curves [$q_E$](*t*) and $p_{i-j}$(*t*) for the protein CI2 at *T/$T_f$=0.94*, the same kind of reconstruction of the hierarchy of events which leads the protein to its native structure has been performed. The curve [$q_E$](*t*) is well fitted by two exponential of characteristic time *30 ns* (*0.3M MCs*) and *1 μs* (*10M MCs*) (data not shown). The first group of contacts which are formed in the first *0.1ns* (*1k MCs*) are all local ones and involves the first turn, essentially all the α−helix and the turn between the strands β4 and β5 (cf. Fig. 7(c)-B see also Fig. 13(I)). Note that the fast folding of the α−helix does not imply strong stability. In fact, as shown in Fig. 7(c)-A, only the two ends of the helix (contacts 13-16, 22-25 and 24-27) are well structured in the unfolded state. The folding proceeds with the formation of the native contacts between strands β4 and β5 and afterwards between strands β4 and β6. This process ends after nearly *20 ns (200k MCs)* when the biggest time gap occur before the next strongly interacting structures, strand β3 and β4, meet each other and dock. This occurs after *0.6-0.95 μs (5-9.5M MCs)* and it is the strongest constraint to the conformational freedom of the chain which closes the reactive site loop (see Fig. 13(III)).

The folding of the CI2 protein (folding time $\tau_f \approx 1$ *μs, 10M MCs*), a trajectory of which is shown in Fig. 11, can also be described in terms of the LES: $S_1((29-34)=β3)$, $S_2((45-52)=β4)$ and $S_3((55-64)=β5+β6)$. Note that this corresponding fragments of the protein used as peptides (3:1 ratio) strongly inhibit the folding of the protein. From the analysis of Fig. 5 one can read the hot and warm amino acids associated with protein CI2. These are (see $\Delta\Delta G_{U-N}$) 29, 47, 49, 50, 57 and 8, 17, 24, 30$^*$, 32, 39, 51, 52$^*$, 58, 60 respectively, the number marked with stars corresponding to warm-hot amino acids. Consequently, the three LES contain all of the hot amino acids and six of the 10 warm amino acids, in particular the warm-hot ones. The formation of the contacts between the $S_2$ and $S_3$ LES (i.e. between β4 and β5+β6) very early in the folding process (*20ns≈200k MCs*, see Fig. 13(II)) leads to a closed LES ($S_{II}(β4+β5+β6)$). Most of the folding time is spent for $S_I(=S_1)$ and $S_{II}$ to explore conformation space to establish the corresponding native contacts. When this happen (*0.6-0.95μs≈6-9.5M MCs*), the folding nucleus is formed (see also Table 4). Shortly after, the protein reaches the native conformation.

DISCUSSION

**Hierarchy of events in protein folding**

The introduction of a residue-dependent pair potential extends the traditional Gō model so as to obtain a better overall better correlation with the experimental data. Consequently, we expect that also the description of the folding events, which are usually not detectable experimentally, is more realistic. For example, simulations of the folding of Protein G made with a standard Gō model show three different folding pathways (Shimada et al. 2002), according to which the protein forms in the intermediate state either the first hairpin, or the second hairpin, or both. Alternatively, our simulations indicate that the second hairpin folds in less than a nanosecond with probability close to 1, while the contacts between strands β1 and β4 are formed only later, corresponding to the post critical FN and thus the transition to the native state. The difference between the two results is due to the fact that in the standard Gō model the folding is only determined by the geometry of the native conformation, which in the case of Protein G is essentially symmetric with respect to the plane normal to the β-sheet. The fact that our modified Gō model introduces different contact energies for the two hairpins breaks this symmetry, allowing the second hairpin to fold faster than the rest of the structures.

The overall picture which emerges for all the three proteins under consideration is a hierarchical folding which is compatible with the experimental data available and which supplements them when this is not available (e.g., on short time scales, etc.). In the case of Protein G, the early events are the formation of the second hairpin and partially of the first (few nanoseconds), and some local contacts in the helix (hundreds of picoseconds). The rate limiting step of the folding process is the formation of the contacts between the two hairpins (milliseconds).

The early formation of these structures, two of which can be viewed as LES, is then crucial for the overall folding of the protein, as already shown in the case of simpler protein models (Broglia et al. 2004, Tiana et al. 2001, Broglia et al. 2001a). Furthermore, their formation is essentially independent on the rest of the protein. In fact, the formation dynamics of these structures display a single-exponential behavior (cf. Fig. 8), compatible with the idea of spontaneous formation. A contact probability $p_{i\text{-}j}(t)$ following a single-exponential dynamics suggests a two-state scenario described by the equation

$$\frac{dp_{ij}}{dt} = a_{ij}[1 - p_{ij}(t)] - b_{ij} p_{ij}(t) ,\tag{4}$$

where the inward and outward rates (i.e., $a_{ij}$ and $b_{ij}$, respectively) are constant. Constant rates imply that the formation of the contact between the $i$th and $j$th residue does not depend on the degree of formation of any other contact (which, in turn, would depend on time). The fact that the dynamics of bonds across the two hairpins is non-exponential (cf. Fig. 8) suggests that the associated rates $a_{ij}$ and $b_{ij}$ depend on time, that is on the degree of formation of the hairpins themselves. These results are in agreement with circular dichroism and NMR spectra of isolated fragments of Protein G in solution. These experiments indicate that the first hairpin is partially structured close to the turn, while the second hairpin is stable (Blanco et al. 1994).

The double exponential dynamics of the overall $[q_E](t)$ displayed in Fig. 8 reflects the two hierarchies of events discussed above, that is formation of local elementary structures (the two hairpins) and their docking. The fact that at neutral pH one can observe only the slower of them reflects the limits of standard experimental techniques. Anyway, the presence of two time scales in the dynamics does not necessarily imply the presence of metastable intermediates populated at equilibrium. In fact, this is not the case for our simulations (cf. Fig. 3) a result which agrees with those of micro calorimetry experiments (Alexander et al. 1992).

The same kind of hierarchical scenario applies for the Src-SH3 protein. Within the first nanoseconds the distal hairpin, the $3_{10}$-helix and the diverging turn get stabilized, while their assembly takes place together with the overall folding of the protein. The results of model calculations agree with the hypotheses done by Baker and co-workers on the basis of φ-value analysis concerning the fact that "the distal hairpin is the most ordered structural element in the transition state", that "the interactions made by the diverging turn residues in the transition state may be greater than indicated by φ-value analysis"[24,] and that "the rate limiting step involves the formation of the distal loop hairpin and the docking of the hairpin onto the diverging turn and the strand following it"[5]. We also find some degree of structure in the RT loop and in the n-src loop, although their formation follow that of the faster structures listed above. The main differences of our results with the interpretation of the experimental data (Grantcharova et al. 1998, Riddle et al. 1999) is the early stabilization of the $3_{10}$ helix and, partially, of the RT loop. On the other hand, unfolding simulations performed by Tsai *et al. (*Tsai et al. 1999) show a late disruption of these two fragments of the protein. Although the unfolding pathway not necessarily is the reverse of the folding one (Zocchi 1997), the results quoted above could indicate that φ-values underestimate the formation of the $3_{10}$ helix and of the RT loop. Note also that the folding mechanism which follows from our analysis essentially agrees with that based on a standard Gō model (Borroguero et al. 2002) and on the implementation of φ-values as harmonic constrains (Lindorff-Larsen et al. 2004), although in the former the RT loop looks less structured than in our simulations.

CI2 is usually taken as example of a protein which folds in a non-hierarchical fashion, following the "nucleation-condensation" mechanism (Karplus et al. 1976). In agreement with the idea of Baldwin and Rose (Baldwin et al. 1999), we show that even in the case of CI2 the folding process is hierarchic. The reason why experiments do not recognize the folding of CI2 as hierarchic is not so much that the LES are not stable as suggested in Baldwin et al. 1999, but that they coincide only marginally with secondary structures which can be detected by typical experimental techniques. In fact, the first folding events are the stabilization of some (but not all) contacts in the α-helix and some contacts between the strands β4-β5 and β4-β6 on nanoseconds time scale. The overall folding takes place when the contacts between strands β3-β4, β2-β5 and β1-β6 get formed. These results agree with the experimental evidence that secondary and tertiary structures appear concurrently (Itzhaki et al. 1995), but this does not mean that the protein does not display early-formed LES which guide the folding process. It only means that they do not coincide with secondary structures. One can identify the LES as a region beyond the N-terminal of the helix and the region enclosed between strands β4-β5. This picture is also in overall agreement with the simulations of Li *et al.* and Clementi *et al.* (Li et al. 2001, Clementi et al. 2000). Note that the description of the folding given above, implies a departure from the chronological sequence of events that one can observe (see *Results* section) to a casual sequence, where the formation of LES is needed to reach fast the post-critical nucleus and eventually the native state. The first argument which supports this departure is that the formation of stable local contacts reduces the entropy of the chain allowing for a faster search through the conformational space of the remaining contacts to form the post-critical folding nucleus. Moreover, if two structures which should assemble together have already their correct shape (i.e. have formed most if not all their internal contacts), once they find each other they can dock with further ado allowing for a fast formation of the corresponding native contacts, as opposed to potential time wasting trapping configurations, configuration which lower both the energy and the entropy at once.

**The unfolded state**

An interesting result of the above simulations is that the unfolded state of the three proteins display some degree of residual structure, corresponding to the local elementary structures which eventually guide the folding process. We are conscious that our description of the unfolded state is biased both because of the neglect of non-native interactions and because the interaction parameters are calculated in the crystallographic conformation. However, the above results agree with a number of direct and indirect evidences. In the case of Src-SH3 NMR studies indicate that the diverging turn is partially formed in the denatured state (Yi et al. 1998). Protein L, which is structurally similar to Protein G but has a markedly different sequence, display in 2M guanidinium a non-random behaviour in the first hairpin (Yi et al. 2000). Combination of NMR and molecular dynamics simulations indicate that the unfolded state of CI2 display some helical structure (Kazmirski et al. 2001).

**Model calculation of the effect of mutations**

The model allows to calculate the effects of mutations on the free energy of the protein following different schemes. The best agreement with the experimental data is found for $\Delta\Delta G_{\ddagger-N}$, the correlation coefficient of the three proteins ranging between *0.57* and *0.65*, and the mean square deviation between *0.48* and *0.86 kcal/mol*. On the contrary, the model estimates of $\Delta\Delta G_{\ddagger-U}$ are quite poor, the correlation coefficient being ≤ *0.2*, and is as poor as the standard Gō model. At the basis of these results is the fact that Gō models in general and the present modified Gō model in paticular, are tailored to describe the interactions in proximity of the native conformation, that is those interactions which build out the highest energy barrier of unfolding.

The free energy difference $\Delta\Delta G_{U-N}$ can be calculated both from equilibrium simulations and as $\Delta\Delta G_{U-\ddagger} - \Delta\Delta G_{\ddagger-N}$. The latter provides slightly better results (the correlation coefficient ranging between *0.45* and *0.62* and a typical standard deviation of *0.85 kcal/mol*), because it does not suffer from equilibrations requirements. Anyway, the correlation coefficient between the $\Delta\Delta G_{U-N}$ calculated with the two methods is ~*0.9*, a result which indicates that the two-state picture of the folding process holds.

The prediction concerning the ϕ-values is rather poor, the correlation coefficients being between *0.48* and *0.15*. This is a consequence of the conspicuous error propagation which is implicit in the definition of ϕ-values. Being a ratio (cf. *Methods*), the relative error $\varepsilon_\phi$ associated with a ϕ-value is $(\varepsilon_{U-\ddagger}^2 + \varepsilon_{U-N}^2)^{1/2}$, where $\varepsilon_{U-\ddagger}$ and $\varepsilon_{U-N}$ are the relative errors which affect the numerator and the denominator, respectively. For example, being *0.85 kcal/mol* the typical error in the prediction of $\Delta\Delta G_{U-N}$, sites with $\Delta\Delta G_{U-N} \approx$ *1 kcal/mol* will display an error larger than 100% on the ϕ-values. Particular affected of this problem are those sites which build out native contacts already in the unfolded state. Since these sites belong to LES, the consequence for the description of the folding process are most important. On the other hand, mutations on these sites rise not only the free energy of the native and of the transition state, but also that of the unfolded state, giving low values of $\Delta\Delta G_{U-N}$.

CONCLUSIONS

We have extended the standard $C_\alpha$–Gō model so as to better account for the chemical diversity of the different types of amino acids forming a protein. Besides calculating the effects of mutations (as $\Delta\Delta G_{U-N}$, $\Delta\Delta G_{\ddagger-N}$ and $\Delta\Delta G_{\ddagger-U}$) and comparing them with experiments, we can also investigate the details of the folding pathways which, as a rule, escape standard experimental techniques, but which strongly qualify the folding abilities of the protein. We observe that proteins G, Src-SH3 and CI2 fold through a hierarchical mechanism whose first step is the formation of LES. We think that this description complements well the common view of protein folding as a chemical reaction through a transition state.

METHODS

We adopted an off-lattice model which simplifies the structure of the protein picturing each amino acid as a hard sphere centred at the position of the $C_\alpha$. The interaction between two residues is given by the contact potential defined in Eq. 1. The attraction between native pairs can be thought as a square-well potential of range *R=7.5Å* whose depth is specific to the native contact considered and given by the matrix element $B_{ij}$. The interaction between non native pairs is repulsive for distances $d_{ij} < R$ and zero for $d_{ij} \geq R$. To avoid overlapping between residues of native pair, we define a hard core distance of 99% of their native distance. Moreover, we assume that residue $i$ interact with residue $i+2$ only through a hard core repulsion of range *3.8Å* and that it does not interact with residue $i+1$ maintaining a fixed distance $d_{i,i+1}$=*3.8Å*. A mutation is defined operatively as a switch of all the native contacts that the mutated residue makes to non-native ones.

Thermodynamical sampling has been performed by means of a Metropolis Monte Carlo algorithm. The kinetic calculations have been performed making use of a dynamic Metropolis algorithm, whose solution has been shown to be equivalent to the solution of the associated Langevin equations (Kikuchi et al. 1991). In order to have an approximated relation between the discrete step and the time measured in seconds, we made 10 independent simulations of protein Src-SH3 observing a linear correlation between the squared mean displacement of the centre of mass and MC time. Comparing this coefficient with a typical diffusing coefficient for a globular protein, we obtained the relation *$10^6$ MC steps ≈ $1 \cdot 10^{-7}$ s*. Note that the reduced degrees of freedom explicitly treated by the model may lead to a reduced viscosity and thus to a shorter scale of times than that one would have obtained had one considered a full atom description of the heteropolymer.

We define operatively the native state for each protein as the set of conformations displaying *$q_E$>0.65* (*$q_E$>0.75* for the Src-SH3) and *dRMSD<5Å*. As a consequence, the unfolded state is characterized by conformations displaying a pair *($q_E$,dRMSD)* outside this interval. We have performed 200 simulations starting from the folded conformation recording the first transition time towards the unfolded state obtaining the transition probability as a function of time $P_{N \to U}(t)$. This curve is well fitted by an exponential whose characteristic time gives the inverse of the unfolding rate: *$\tau = 1/k_u$*. In the same way, the folding rate $k_f$ is obtained from the exponential fit of $P_{U \to N}(t)$ resulting from 200 simulations starting from random generated conformations. These two sets of simulations are repeated for each mutation and for each protein. The temperatures of the samplings are fixed for each protein to *$T/T_f$=0.94* for Protein G, *$T/T_f$=1.1* for Src-SH3 and *$T/T_f$=1.05* for CI2.

From the kinetic rates of folding and unfolding of the wild-type and mutated protein, we have calculated the differences of free energies between the transition and the unfolded state through $\Delta\Delta G_{\ddagger-U} = T \log k_f^{mut}/k_f^{wt}$ and between the transition and the native state through $\Delta\Delta G_{\ddagger-N} = T \log k_u^{mut}/k_u^{wt}$. The variations of free energy between native and unfolded state $\Delta\Delta G_{U-N}$ are calculated in two different ways: from equilibrium simulations (20 simulations of *50 μs* per mutation per protein) through $\Delta\Delta G_{U-N}^{(eq)} = T \log P_U^{mut}/P_U^{wt} - T \log P_n^{mut}/P_N^{wt}$, where $P_N$ is the probability for the protein to be in the native state and $P_U=1-P_N$, and from the kinetic rates through $\Delta\Delta G_{U-N}^{(kin)} = T \log k_u^{mut}/k_u^{wt} - T \log k_f^{mut}/k_f^{wt}$. The φ-values are calculated either from $\phi = -\Delta\Delta G_{\ddagger-U}/\Delta\Delta G_{U-N}$ or from $\phi = 1 - \Delta\Delta G_{\ddagger-N}/\Delta\Delta G_{U-N}$.

The pdb code for the Protein G, Src-SH3 and CI2 used in the present paper are respectively 1PGB, 1FMK, 2CI2.

## TABLES

|  | $\Delta\Delta G_{U-N}^{exp}$ | | $\Delta\Delta G_{\ddagger-U}^{exp}$ | $\Delta\Delta G_{\ddagger-N}^{exp}$ | $\phi^{exp}$ | |
|---|---|---|---|---|---|---|
|  | $\Delta\Delta G_{U-N}^{(eq)}$ | $\Delta\Delta G_{U-N}^{(kin)}$ | $\Delta\Delta G_{\ddagger-U}$ | $\Delta\Delta G_{\ddagger-N}$ | $\phi^{(1)}$ | $\phi^{(2)}$ |
| Protein G | 0.41 (0.90) | 0.50 (0.85) | 0.21 (0.41) | 0.64 (0.65) | 0.36 (0.27) | 0.48 (0.44) |
| Protein G (standard Gō model) | -- | 0.41 (0.90) | 0.21 (0.41) | 0.47 (0.75) | -- | 0.46 (0.69) |
| Src-SH3 | 0.34 (0.64) | 0.45 (0.60) | 0.01 (0.52) | 0.57 (0.48) | 0.13 (0.36) | 0.19 (0.51) |
| CI2 | 0.65 (0.97) | 0.62 (1.00) | 0.00 (0.45) | 0.58 (0.86) | 0.15 (0.26) | 0.00 (0.63) |

Table 1: The correlation coefficient $r$ (and, in parenthesis, the associated root mean square deviation ($\sigma$)) between the experimental and calculated mutation parameters (cf. *Methods*) for the three proteins. In the second row the correlations and the rmsd are calculated between the experimental and the theoretical values obtained for protein G with a standard Gō model. φ-values are obtained from: $\phi^{(1)} = -\Delta\Delta G_{\ddagger-U}/\Delta\Delta G_{U-N}^{(eq)}$; $\phi^{(2)} = 1 - \Delta\Delta G_{\ddagger-N}/\Delta\Delta G_{U-N}^{exp}$.

| $S_1$ (4–9) | β1 (1–9) | | (post critical) FN | FPT |
|---|---|---|---|---|
| $S_2$ (41–46) | β3 (43–46) | $S_{II}$ (41–54) | | |
| $S_3$ (49–54) | β4 (50–56) | β3+β4 Turn4 (47–49) | | |
| | 0.1–1.0ns ($10^3$–$10^4$ MCs) | | 1.0–1.5μs (10–15M MCs) | 2μs (20M MCs) |

Table 2: Protein G LES and (in parenthesis) initial and final amino acids number forming them. The formation of the native contacts between $S_2$ and $S_3$ leads to the closed $S_{II}(41-54)$ LES (and also to turn 4), while the docking of $S_1$ and $S_{II}$ gives rise to the FN.

| $S_1$ (3–10) | β1 (2–6) | RT–loop (8–19) | | |
| | 1/4 RT–loop (8,9,10) | 30–50ns | (post critical) FN | FPT |
| $S_2$ (18–26) | 1/6 RT–loop (18,19) +3/5 β2 (24,25,26) | diverging turn (20–27) | | |
| $S_3$ (36–44) | β3 (36–41) | distal hairpin (36–51) | | |
| $S_4$ (47–51) | β4 (47–51) | | | |
| 0.1ns | | 5–10ns | 1.6–1.8μs (16–18M MCs) | 2μs (20M MCs) |

Table 3: Protein Src-SH3 LES and (in parenthesis) initial and final amino acids number forming them. The formation of the native contacts between $S_3$-$S_4$ gives rise to the distal hairpin (which can be viewed as a closed LES), while that corresponding to $S_1$-$S_2$ gives rise to the formation of the RT-loop and of the diverging turn (second closed LES). The docking of the two closed LES gives rise to the (post critical) folding nucleus (FN).

| $S_1$ (29–34) | β3 (28–34) | | | |
| $S_2$ (45–52) | β4 (45–52) | $S_{II}$ (45–64) | (post critical) FN | FPT |
| $S_3$ (55–64) | β5+β6 | | | |
| | 20ns(=200k MCs) | 0.6–0.95μs (6–9.5M MCs) | 1μs (10M MCs) | |

Table 4: Protein CI2 LES and (in parenthesis) initial and final amino acids number forming them. The formation of the native contacts between $S_2$-$S_3$ gives rise to a closed LES $S_{II}$ while the docking of $S_I(=S_1)$ and $S_{II}$ leads to the FN.

# REFERENCES

Abkevich, V.I., Gutin, A.M. and Shakhnovich, E.I. 1994. Specific nucleus as the transition state for protein folding. *Biochemistry* **33:** 10026-10031.

Alexander, P., Fahnestock S., Lee, T., Orban, J. and Bryan, P. 1992. Thermodynamic analysis of the folding of the streptococcal Protein G IgG-binding domains B1 and B2. *Biochem.* **31:** 3597-3603

Baldwin, R. L. and Rose, G. D. 1999. Is protein folding hierarchic? *TIBS* **24:** 26-33.

Berendsen, H.J.C., van der Spoel, D. and van Drunen, R. 1995. GROMACS: A message--passing parallel molecular dynamics implementation. *Comp. Phys. Comm.* **91:** 43-55.

Blanco, F. G., Rivas, G. and Serrano, L. 1994. Folding of protein G B1 domain studied by conformational characterization of fragments comprising its secondary structure. *Eur. J. Biochem.* **230:** 634-649.

Borroguero, J.M., Dokholyan, N.V., Buldryev, S.V., Shakhnovich, E.I. and Stanley, H.E. 2002. Thermodynamics and folding kinetics analysis of the SH3 domain from discrete molecular dynamics. *J. Mol. Biol.* **318:** 863-876.

Broglia, R.A. and Tiana, G. 2001a. Hierarchy of events in the folding of model proteins. *J. Chem. Phys.* **114:** 7267-72.

Broglia, R.A. and Tiana, G. 2001b. Reading the three-dimensional structure of a protein from its amino acid sequence. *Proteins* **45:** 421-427.

Broglia, R.A., Tiana, G. and Provasi, D. 2004. Simple model of protein folding and of non-conventional drug design. *J. Phys. Cond. Mat.* R **16:** R111-R144.

Broglia, R.A., Tiana, G., Sutto, L., Provasi, D. and Simona, F. 2005. Design of HIV-1-PR inhibitors that do not create resistance: Blocking the folding of single monomers. *Protein Science* **14:** 2668-2681.

Bryngelson, J. D. and Wolynes, P.G. 1987. Spin glasses and the statistical

mechanics of protein folding. *Proc. Natl. Acad. Sci. USA* **84:** 7524-7528.

Clementi, C., Nymeyer H. and Onuchic J.N. 2000. Topological and energetic factors: what determines the structural details of the transition state ensemble and en-route intermediates for protein folding? *J. Mol. Biol.* **298:** 937-953.

Gō, N. 1983. Theoretical studies of protein folding. *Annu. Rev. Biophys. Bioengin.* **12:** 183-210.

Grantcharova, V., Riddle, D., Santiago, J. and Baker, D. 1998. Important role of hydrogen bonds in the structurally polarized transition state for folding of the src SH3 domain. *Nature Struct. Biol.* **5:** 714-720.

Guerois, R., Nielsen, J. E. and Serrano, L. 2002. Predicting Changes in the Stability of Proteins and Protein Complexes: A Study of More Than 1000 Mutations. *J. Mol. Biol.* **320:** 369-387.

Hansen, A., Jensen, M.H., Sneppen, K. and Zocchi, G. 1998. A hierarchical scheme for cooperativity and folding in proteins. *Physica A* **250:** 335-351.

Humphrey, W., Dalke, A. and Schulten, K. 1996. VMD – Visual Molecular Dynamics. *J. Molec. Graphics* **14.1:** 33-38.

Itzhaki, L. S., Otzen, D. E. and Fersht, A.R. 1995. The structure of the transition state of chymotrypsin inhibitor 2 analysed by protein enguineering methods:evidence for a nucleation-condensation mechanism for protein folding. *J. Mol. Biol.* **254:** 260-288.

Karplus, M. and Weaver, D. L. 1976. Protein folding dynamics. *Nature*, **260:** 404-406.

Kazmirski, S. L., Wong, K. B., Freund, S.M.V., Tan., Y. J., Fersht, A. R and Daggett V. 2001. Protein folding from a highly disordered denatured state: folding pathway of chymotrypsin inhibitor 2 at atomic resolution. *Proc. Natl. Acad. Sci. USA* **98:** 4349-4354.

Khan, F., Chuang, J.I., Gianni, S. and Fersht, A.R. 2003. The Kinetic Pathway of Folding of Barnase. *J. Mol. Biol.* **333:** 169-186.

Kikuchi, K., Yoshida, M., Maekawa, T. and Watanbe, H. 1991. Metropolis Monte Carlo method as a numerical technique to solve Fokker-Planck equation. *Chem. Phys. Lett.* **185:** 335-338.

Lazaridis, T. and Karplus, M. 1999. Discrimination of the native from misfolded protein models with an energy function including implicit solvation. *J. Mol. Biol.* **288:** 477-487.

Lesk, A. M. and Rose, G. D. 1981. Folding units in globular proteins. *Proc. Natl. Acad. Sci. USA*. **78:** 4304-4308.

Levy, Y., Caflish, A., Onuchic, J.N. and Wolynes, P.G. 2004. The folding and dimerization of HIV-1-Protease: Evidence for a stable monomer from simulations. *J. Mol. Biol.* **340:** 67-69.

Li, L and Shakhnovich, E.I. 2001. Constructing, verifying and dissecting the folding transition state of CI2 with all atom simulations. *Proc. Natl. Acad. Sci. USA* **98:** 13014-13018.

Lindorff-Larsen, K., Vendruscolo, M., Paci, E. and Dobson, C. M. 2004. Transition states for protein folding have native topologies despite high structural variability. *Nature Struct. Biol.* **11:** 443-449.

Martinez, J. C., Pisabarro, M.T. and Serrano, L. 1998. Obligatory steps in protein folding and the conformational diversity of the transition state. *Nature Struct. Biol.* **5:** 721-729.

McCallister, E. L., Alm, E. and Baker, D. 2000. Critical role of beta-hairpin formation in Protein G folding. *Nature Struct. Biol.* **7:** 669-763.

Panchenko, A.R., Luthey-Schulten, Z. and Wolynes, P.G. 1995. Foldons, protein structural modules, and exons. *Proc. Natl. Acad. Sci. USA* **93:** 2008-2013.

Park, S.H., Ramachandra Shastry, M. C. and Roder, H. 1999. Folding dynamics of the B1 domain of protein G explored by ultrarapid mixing. *Nature Struct. Biol.* **6:** 943-947.

Pearlman, D.A., Case, D.A., Caldwell, J.W., Ross, W.S., Cheatham III, T.E., DeBolt, T., Ferguson, D., Seibel, G. and Kollman P. 1995. AMBER, a package of computer programs for applying molecular mechanics, normal mode analysis, molecular dynamics and free energy calculations to simulate the structural and energetic properties of molecules. *Comp. Phys. Commun.* **91:** 1-41.

Ptitsyn, O. B. and Rashin, A. A. 1975. *Biophys. Chem.* **3:** 1-20.

Religa, T.L., Marhson, J.S., Mayor, U., Fremd S.M.V., and Fersht A.R. 2005. Solution Structure of a protein denatured state and folding intermediate. *Nature* **437:** 1053-1056.

Riddle, D., Grantcharova, S.V.P., Santiago, J.V., Alm, E., Ruczinski, I. & Baker, D. 1999. Experiment and theory highlight role of native state topology in SH3 folding. *Nature Struct. Biol.* **6:** 1016-1024.

Shimada, J. and Shakhnovich, E. I. 2002. The ensemble folding kinetics of protein G from an all-atom Monte Carlo simulation. *Proc. Natl. Acad. Sci. USA*, **99:** 11175-11180.

Tiana, G. and Broglia, R.A. 2001. Statistical analysis of contact formation in the folding of model proteins. *J. Chem. Phys.* **108:** 2503-2510.

Tsai, J., Levitt, M. and Baker, D. 1999. Hierarchy of structure loss in MD simulations of src SH3 domain unfolding. *J. Mol. Biol.* **291:** 215-225.

Yi, Q., Bystroff, C. and Baker, D. 1998. Prediction and structure characterization of an independently folding substructure in the src SH3 domain. *J Mol Biol.* **283:** 293-300.

Yi, Q., Scalley-Kim, M. L., Alm, E. J. and Bajer, D. 2000. NMR characterization of residual structure in the denatured state of protein L. *J. Mol. Biol.* **299:** 1341-1351.

Zocchi, G. 1997. Proteins unfold in steps. *Proc. Natl. Acad. Sci. USA* **94:** 10647-10651.

FIGURES

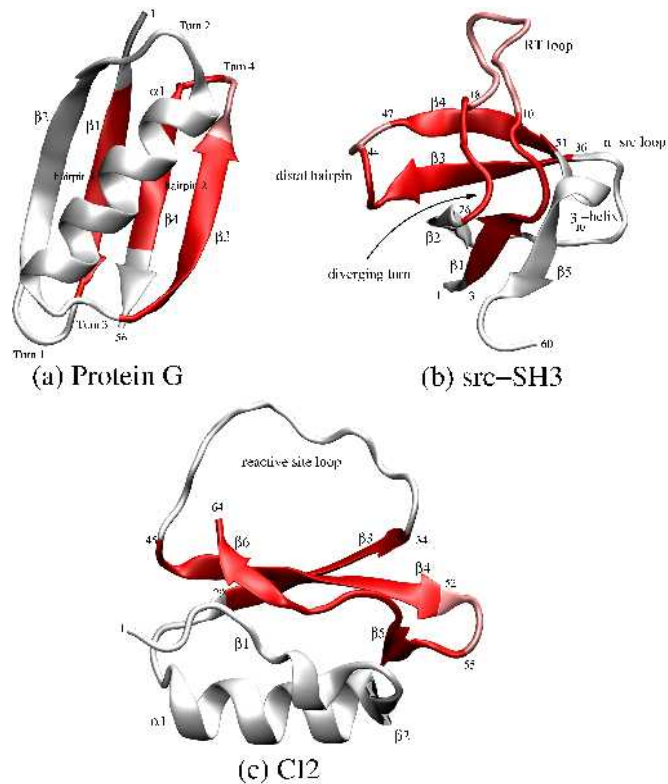

Figure 1: A cartoon representation of proteins G, Src-SH3 and CI2 (images created using VMD software (Humphrey et al. 1996)). Highlighted in red are the fragment of the protein corresponding to the LES of the protein and thus containing, in average, the residues with the most stable and early forming contacts.

(a) *Protein G*: It is formed by 56 residues arranged in four β-motifs (β1(1-9), β2(12-20), β3(43-46) and β4(50-56)) an α-helix (α1(23-36)) and four turns (Turn1(10,11), Turn2(21,22), Turn3(37-41) and Turn4(47-49)). The LES are $S_1$(4-9), $S_2$(41-46) and $S_3$(49-54). They contain essentially all of the hot amino acids (26, 41, 45, 52, 54, see Fig. 4) and about half of the warm amino acids (6, 7, 20, 31, 34, 35, 51, 53, see Fig. 4). The docking $S_2$-$S_3$ (essentially equivalent to the docking of β3-β4) leads to a closed LES (see text and Table 2, as well as Broglia et al. 2001b). We have thus also coloured Turn4, although strictly speaking it does not belong to any of the LES.

(b) *Protein Src-SH3*: It is formed by 60 residues arranged as follows; β1(2-6), RT-loop(8-19), diverging turn(20-27), β2(24-28), β3(36-41), distal hairpin(36-51), β4(47-51), $3_{10}$-helix(51-54) and β5(54-57). The LES are $S_1$(3-10), $S_2$(18-26), $S_3$(36-44) and $S_4$(47-51). They contain all of the hot (10, 18, 20, 26, 50) and warm (5, 7, 23, 24, 38, 41, 44*, 48*, 49, 51; asterisk warm-hot) amino acids (see Fig. 5). The docking of $S_3$-$S_4$

LES gives rise to a closed LES and stabilizes the distal hairpin which. Even if all of its amino acids do not belong to $S_3+S_4$ (i.e. 42-46) we have coloured the whole motif, (clearer tone used for amino acids outside $S_3+S_4$). The docking of $S_1$-$S_2$ gives rise to a second (closed) LES leading also to the formation of the RT-loop and the diverging turn (see Table 3). We have thus also coloured the fragment 11-17 (with a clearer tone) although this fragment of the protein does not strictly belong to any of the LES.

(c) *Protein CI2*: It is formed by 64 residues arranged in the following motifs: β1(3-5), β2(10-11), α1(12-24), β3(28-34), reactive loop (35-44), β4(45-52), β5(55-58) and β6(60-64). The LES are $S_1$(29-34), $S_2$(45-52) and $S_3$(55-64). They contain all the hot amino acids of the protein (29, 47, 49, 50, 57; see Fig.6) and most of the warm amino acids (30, 32, 34, 51, 52, 58, 60; see Fig.6) The docking of $S_2$-$S_3$ gives rise to a (closed) LES. This is the reason why we have also coloured (clearer tone) amino acids 53 and 54 (Turn2), although they actually do not belong to any of the LES.

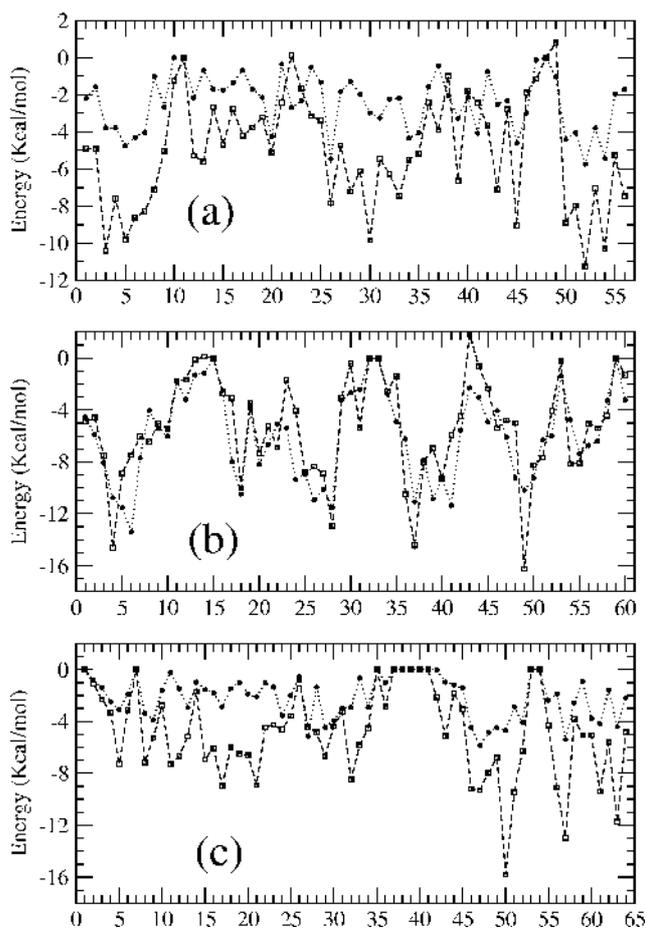

Figure 2: Energies $\Sigma_j B_{ij}$ (solid dots) per monomer in the native conformation of protein G (a), Src-SH3 (b) and CI2 (c), determined making use of the interaction energies $B_{ij}$ calculated from the experimental values of $\Delta\Delta G_{U\text{-}N}$ (see text), in comparison with the corresponding quantities calculated using the software GROMACS (open squares).

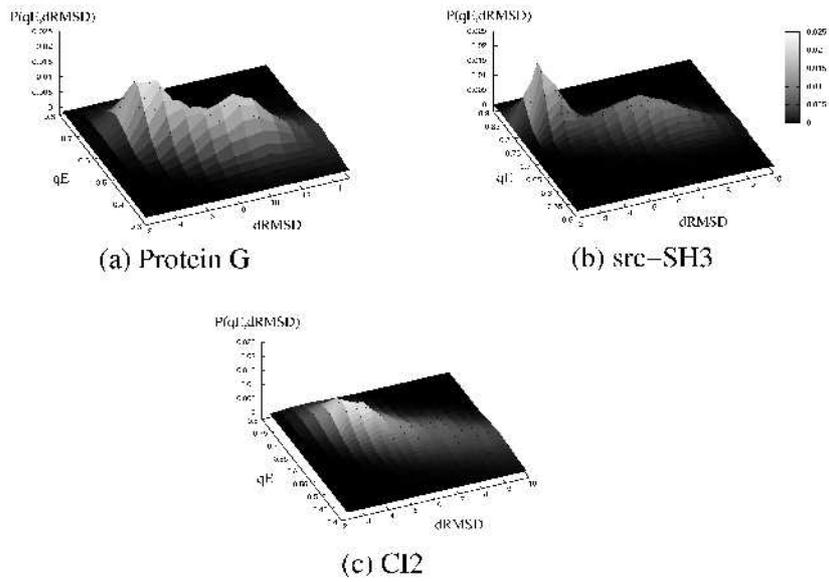

Figure 3: The equilibrium probability of (a) Protein G (at $T/T_f=1.13$), (b) Src-SH3 (at $T/T_f=1.1$) and (c) CI2 (at $T/T_f=1.05$) as a function of the relative energy parameter $q_E$ and of the *dRMSD*. While in the Protein G and the Src-SH3 the two peaks representing the native and the unfolded state, are well defined, for the CI2 the peaks are smoother indicating a less cooperative transition between states.

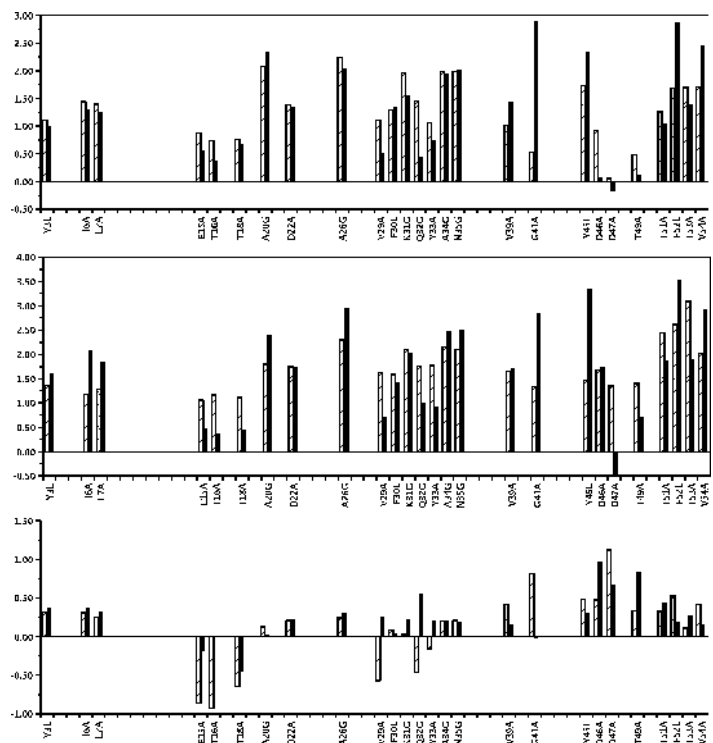

Figure 4: $\Delta\Delta G_{\ddagger\text{-}N}$, $\Delta\Delta G_{U\text{-}N}^{(kin)}$ (*kcal/mol*) and $\phi=1-\Delta\Delta G_{\ddagger\text{-}N}/\Delta\Delta G_{U\text{-}N}^{\exp}$ values associated with protein G. Black histograms correspond to experimental values of the different quantities, while dashed histograms correspond to the prediction of the model.

Figure 5: The same as of Fig. 4, calculated for Src-SH3: from above: the $\Delta\Delta G_{\ddagger\text{-}N}$, the $\Delta\Delta G_{U\text{-}N}^{(kin)}$ in *kcal/mol* and the ɸ-values calculated from ɸ = $1 - \Delta\Delta G_{\ddagger\text{-}N} / \Delta\Delta G_{U\text{-}N}^{exp}$.

Figure 6: The same as of Fig. 4, calculated for CI2: from above: the $\Delta\Delta G_{\ddagger\text{-}N}$, the $\Delta\Delta G_{U\text{-}N}^{(kin)}$ in *kcal/mol* and the $\phi$-values calculated from $\phi = 1 - \Delta\Delta G_{\ddagger\text{-}N} / \Delta\Delta G_{U\text{-}N}^{\exp}$.

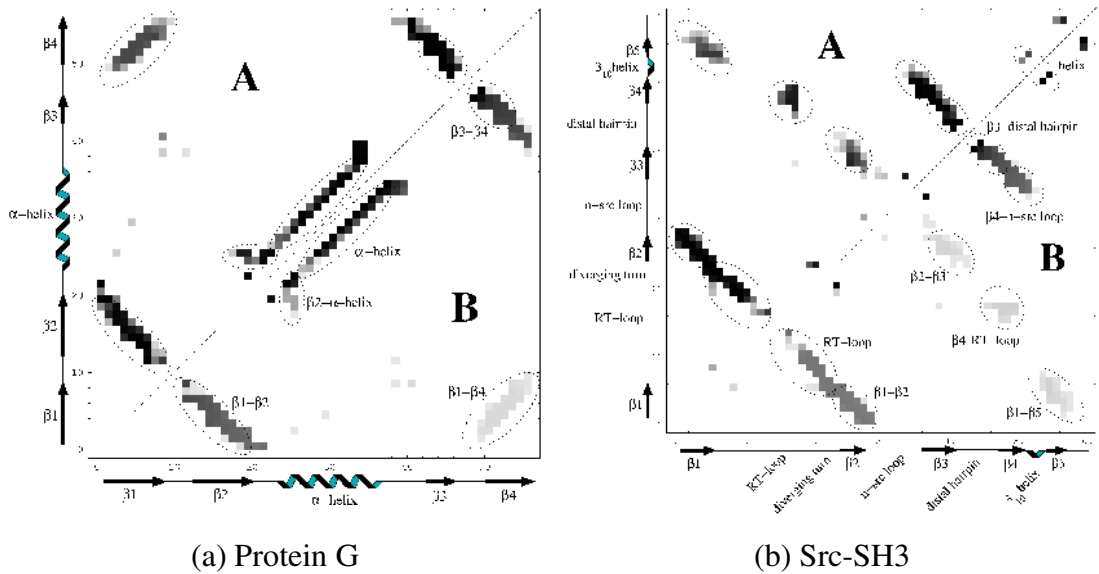

(a) Protein G    (b) Src-SH3

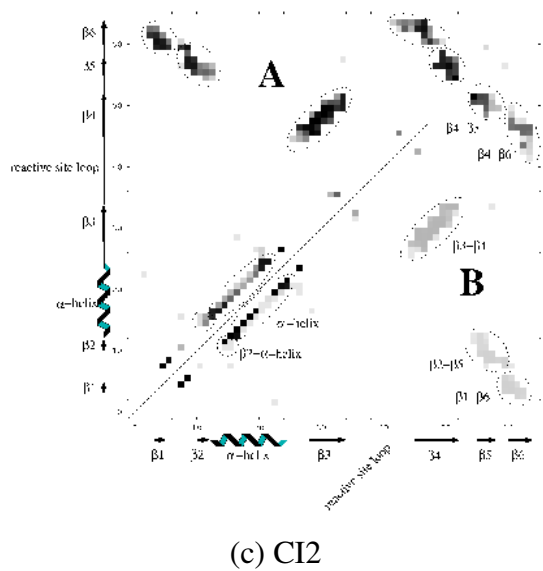

(c) CI2

Figure 7: The contact map of (a) Protein G, (b) Src-SH3 and (c) CI2 at $T<T_f$. The colours qualitatively indicate, in the upper half of the map (labelled (A)) the equilibrium probability of contact formation (black corresponding to the maximum contact stability), while in the lower half (B) the formation times of the contacts are displayed (black squares indicates a formation time of *0.1 ns*, grey squares of *10 ns* and light grey of *1 μs*).

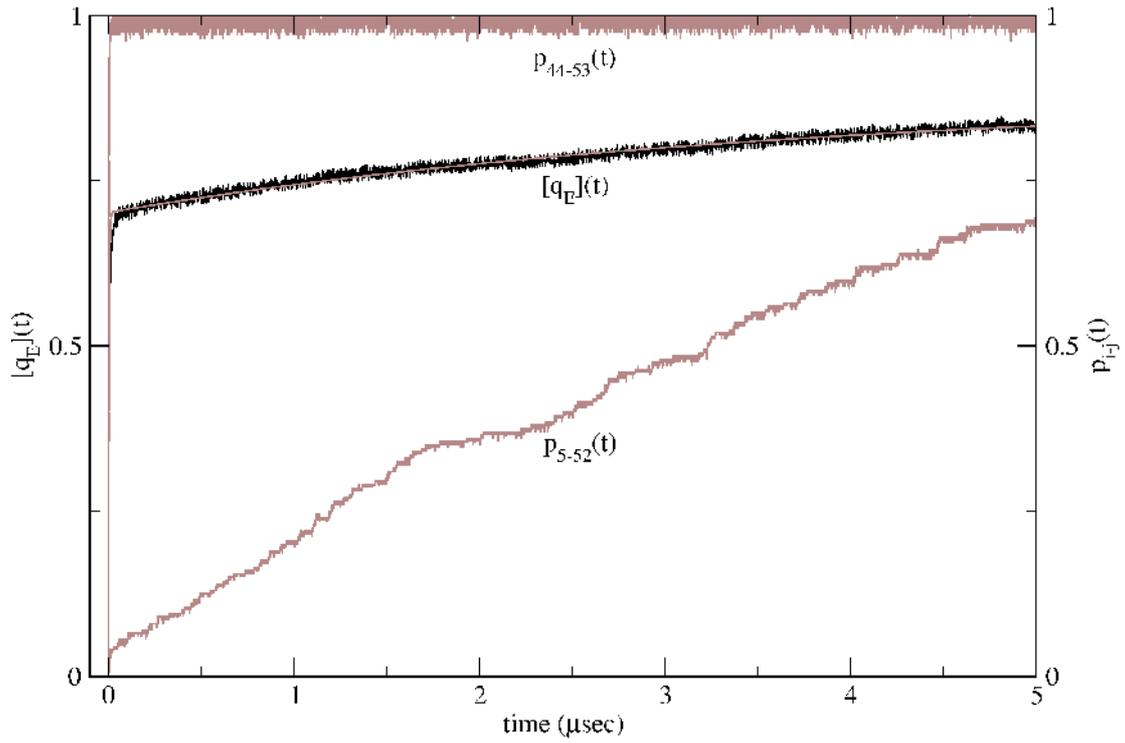

Figure 8: Average similarity parameter $[q_E](t)$ as a function of time, which characterizes the folding dynamics of Protein G at a temperature $T/T_f=0.54$ (black curve) with its exponential fit (grey central curve). In dark grey the formation probability $p_{i-j}(t)$ of contacts 44-53 (fast forming) and 5-52 (slow forming).

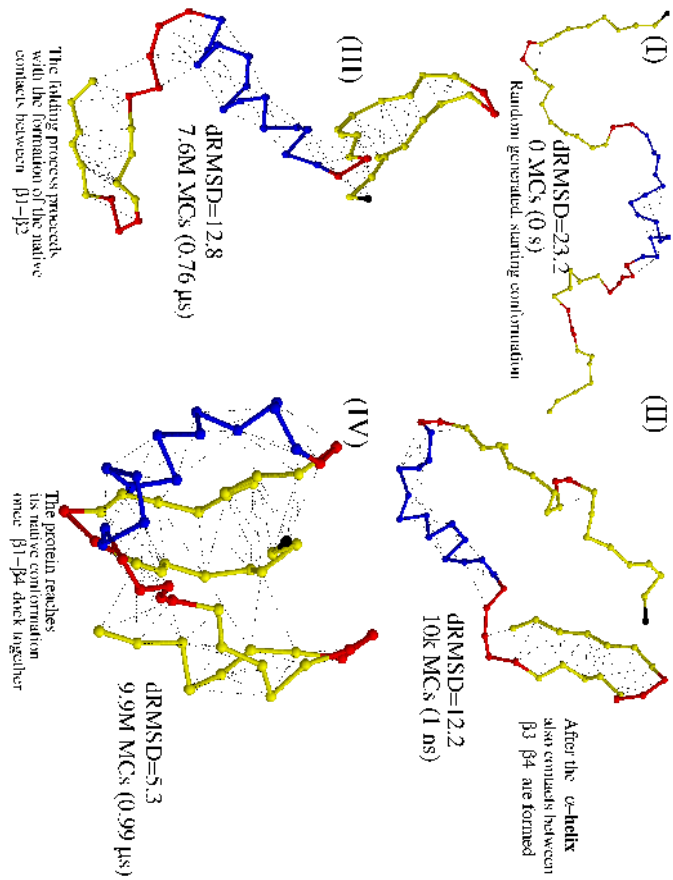

Figure 9: Contact formation diagram for Protein G. The black bead indicates the first residue.

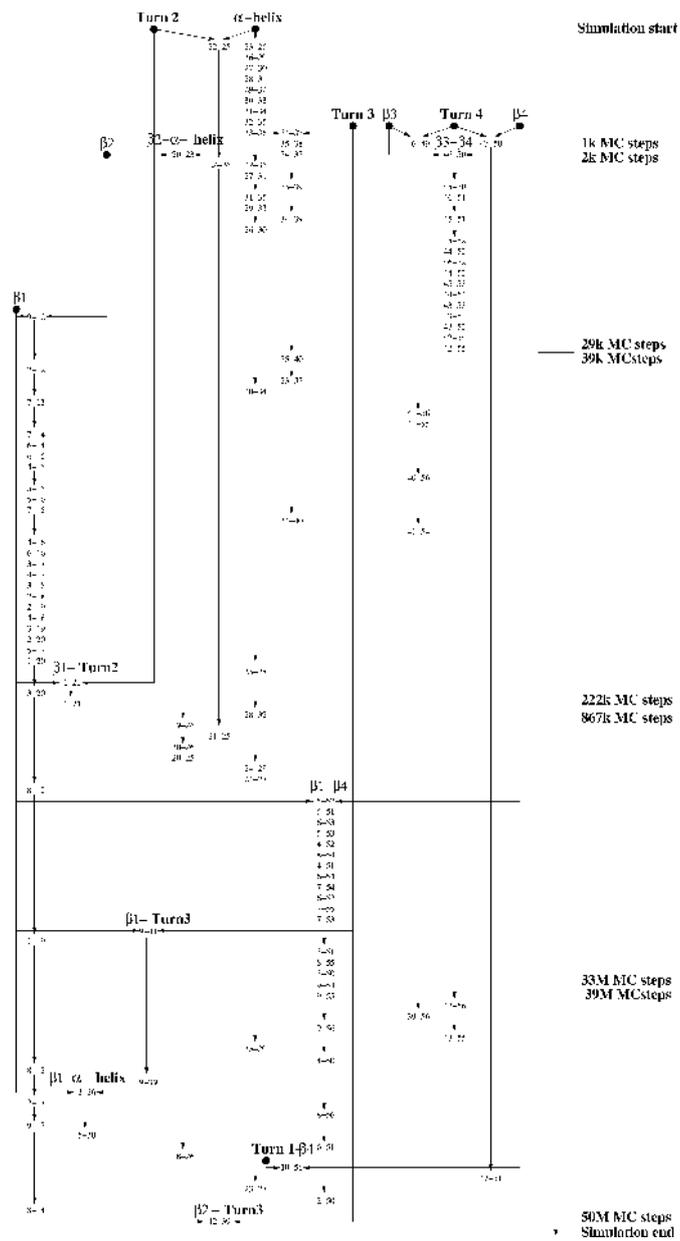

Figure 10: Contact formation flow for Protein G

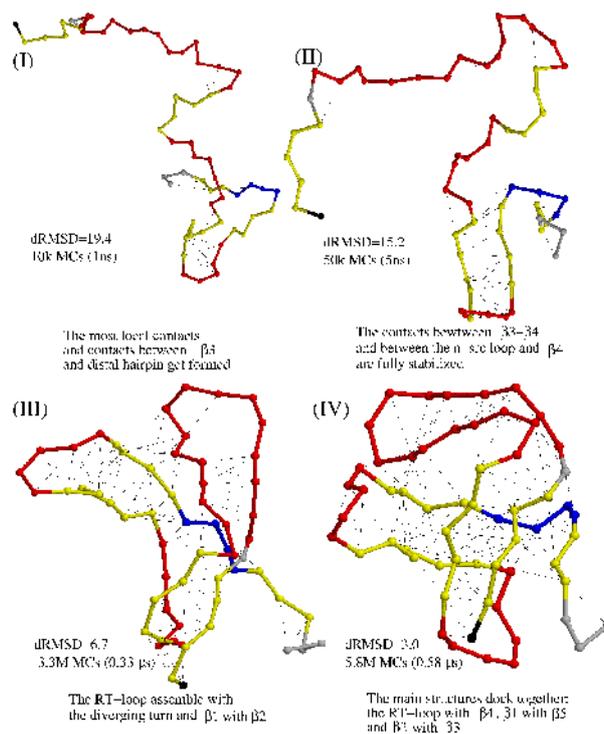

Figure 11: Contact formation diagram for Src-SH3. The black bead indicates the first residue.

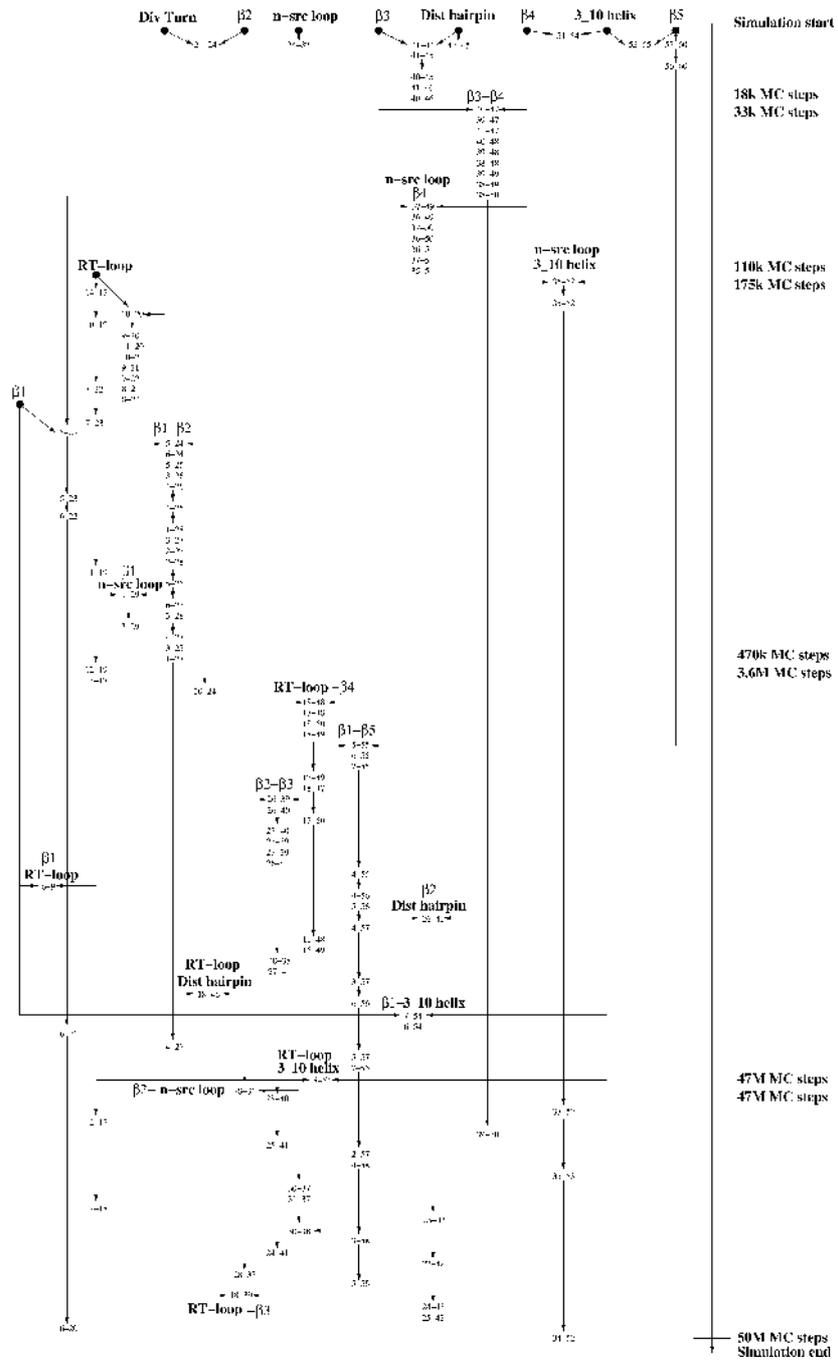

Figure 12: Contact formation flow for Src-SH3

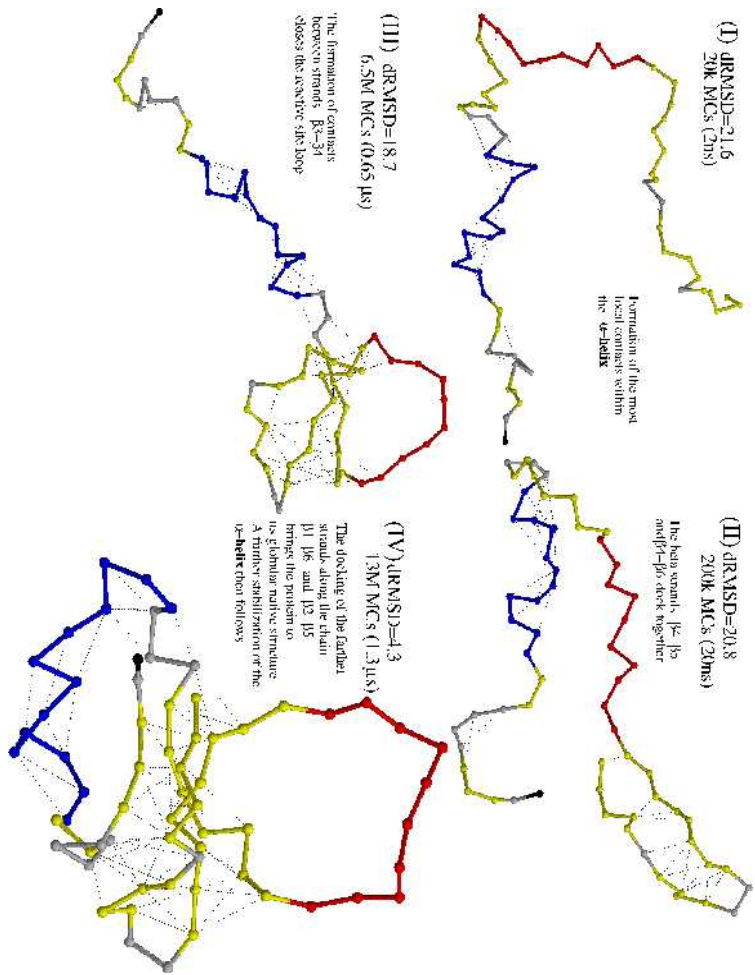

Figure 13: Contact formation diagram for CI2. The black bead indicates the first residue.

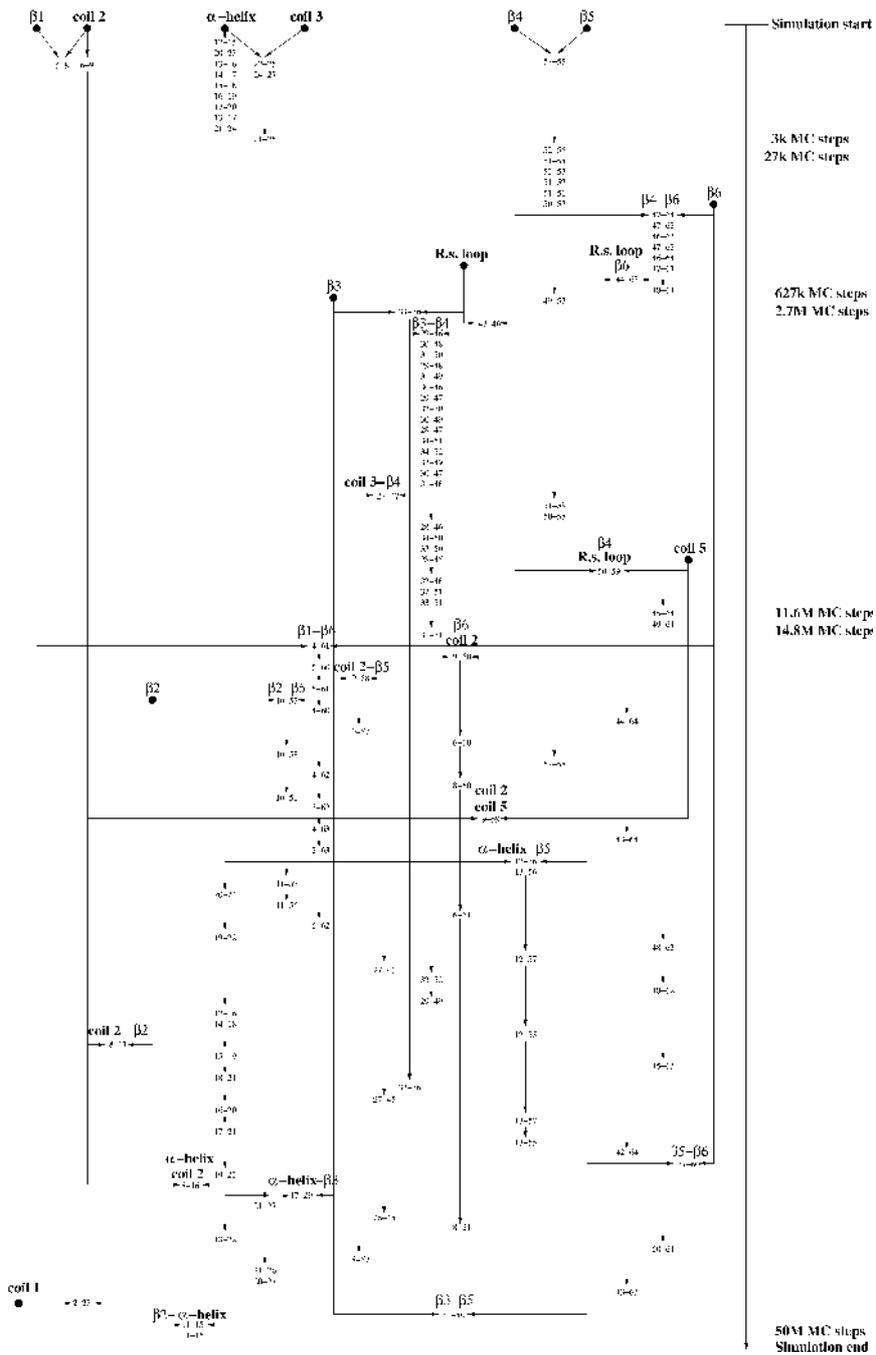

Figure 14: Contact formation flow for CI2